\documentclass[twocolumn]{aastex62}

\usepackage{graphicx}	% Including figure files
\usepackage{amsmath}	% Advanced maths commands
\usepackage{amssymb}	% Extra maths symbols
\usepackage{float}
\usepackage{hyperref}
\usepackage{algorithm}
\usepackage[noend]{algpseudocode}
\usepackage[FIGTOPCAP]{subfigure}
\usepackage{tabularx}

\newcommand{\vect}[1] {\mbox{\boldmath ${#1}$}}

\def\btheta{\boldsymbol{\theta}}
\def\ln{\mathrm{ln}}

\graphicspath{{./}{figures/}}

\shorttitle{Emulating Stellar Population Synthesis}
\shortauthors{Alsing et al.}

\begin{document}

\title{\textsc{speculator}: Emulating stellar population synthesis for fast and accurate galaxy spectra and photometry}

\correspondingauthor{Justin Alsing}
\email{justin.alsing@fysik.su.se}

\author[0000-0003-4618-3546]{Justin Alsing}
\affiliation{Oskar Klein Centre for Cosmoparticle Physics, Department of Physics, Stockholm University, Stockholm SE-106 91, Sweden}

\author{Hiranya Peiris}
\affiliation{Department of Physics and Astronomy, University College London, Gower Street, London, WC1E 6BT, UK}
\affiliation{Oskar Klein Centre for Cosmoparticle Physics, Department of Physics, Stockholm University, Stockholm SE-106 91, Sweden}

\author{Joel Leja}
\altaffiliation{NSF Fellow}
\affiliation{Harvard-Smithsonian Center for Astrophysics, 60 Garden St. Cambridge, MA 02138, USA}

\author{ChangHoon Hahn}
\affiliation{Lawrence Berkeley National Laboratory, 1 Cyclotron Rd, Berkeley CA 94720, USA}
\affiliation{Berkeley Center for Cosmological Physics, University of California, Berkeley, CA 94720, USA}

\author{Rita Tojeiro}
\affiliation{School of Physics and Astronomy, University of St Andrews, North Haugh, St Andrews, KY16 9SS, UK}

\author{Daniel Mortlock}
\affiliation{Department of Physics, Imperial College London, Blackett Laboratory, Prince Consort
Road, London SW7 2AZ, UK}
\affiliation{Oskar Klein Centre for Cosmoparticle Physics, Department of Physics, Stockholm University, Stockholm SE-106 91, Sweden}

\author{Boris Leistedt}
\altaffiliation{NASA Einstein Fellow}
\affiliation{Center for Cosmology and Particle Physics, Department
of Physics, New York University, New York, NY}

\author{Benjamin D. Johnson}
\affiliation{Harvard-Smithsonian Center for Astrophysics, 60 Garden St. Cambridge, MA 02138, USA}

\author{Charlie Conroy}
\affiliation{Harvard-Smithsonian Center for Astrophysics, 60 Garden St. Cambridge, MA 02138, USA}

\begin{abstract}
We present \textsc{speculator} -- a fast, accurate, and flexible framework for emulating stellar population synthesis (SPS) models for predicting galaxy spectra and photometry. For emulating spectra, we use principal component analysis to construct a set of basis functions, and neural networks to learn the basis coefficients as a function of the SPS model parameters. For photometry, we parameterize the magnitudes (for the filters of interest) as a function of SPS parameters by a neural network. The resulting emulators are able to predict spectra and photometry under both simple and complicated SPS model parameterizations to percent-level accuracy, giving a factor of $10^3$--$10^4$ speed up over direct SPS computation. They have readily-computable derivatives, making them amenable to gradient-based inference and optimization methods. The emulators are also straightforward to call from a GPU, giving an additional order-of-magnitude speed-up. Rapid SPS computations delivered by emulation offers a massive reduction in the computational resources required to infer the physical properties of galaxies from observed spectra or photometry and simulate galaxy populations under SPS models, whilst maintaining the accuracy required for a range of applications.% We anticipate that fast SPS emulation will play a key role in future analysis and simulation of galaxy populations with SPS.
\end{abstract}


\section{Introduction}
\label{sec:intro}
Inferring the physical properties of galaxies from observations of the spectral energy distribution (SED) of their emitted light is one of the cornerstones of modern extragalactic astronomy. At the heart of this endeavor is stellar population synthesis (SPS): predictive models for galaxy SEDs that fold together the initial stellar mass function, star formation and metallicity enrichment histories, stellar evolution calculations and stellar spectral libraries, phenomenological dust and gas models, black hole activity etc., to predict the spectrum of a galaxy given some input physical parameters associated with each model component. SPS modeling has a rich history, with a plethora of parameterizations of varying complexity available (see \citealp{conroy2013} and references therein). 

The computational bottleneck in both inferring galaxy properties from observations and simulating catalogs under SPS models, is running the SPS models themselves. Forward-simulating upcoming Stage IV galaxy surveys will demand $\sim10^{10}$ SPS evaluations per catalog simulation. For data analysis, inferring\footnote{e.g., Markov Chain Monte Carlo sampling.} of order ten SPS model parameters for a single galaxy (given some photometric or spectroscopic data) typically requires $\sim10^5-10^6$ SPS model evaluations. If inference is then to be performed for a large sample of galaxies, the number of SPS evaluations and associated computational demands quickly become prohibitive. For recent context, \citet{leja2019} analyzed $\sim 6\cdot 10^4$ galaxies under a 14-parameter SPS model, with a total cost of $1.5$ million CPU hours\footnote{For added context, the CPU time for the \citet{leja2019} analysis would cost around twenty-thousand USD from Amazon Web Services (estimated in 2019).}. With upcoming surveys such as the Dark Energy Spectroscopic Instrument (DESI; \citealp{levi2013,aghamousa2016,aghamousa2018}) posing the challenge of analyzing millions of galaxy spectra, the need to address the bottleneck posed by SPS is clear and urgent.

There are two principal ways of reducing the cost of inference and simulation under SPS models: speeding up individual SPS computations, and (in the case of inference) reducing the number of SPS computations required to obtain robust inferences per galaxy. In this paper we present neural network emulators for SPS spectra and photometry that gain leverage on both fronts. For galaxy spectra, our emulation framework uses principal component analysis (PCA) to construct a basis for galaxy SEDs, and then trains a neural network on a set of generated SPS spectra to learn the PCA basis coefficients as a function of the SPS model parameters. For photometry, we train a neural network to learn the magnitudes directly (for some set of band passes) as a function of the SPS parameters. The result in both cases is a compact neural network representation of the SPS model that is both fast to evaluate, accurate, and has analytic and readily-computable derivatives, thus making it amenable to efficient gradient-based optimization and inference methods (e.g., Hamiltonian Monte Carlo sampling). Furthermore, calling the emulators from a GPU is straightforward, enabling an additional order-of-magnitude speed-up when evaluating many SPS models in parallel.

We demonstrate and validate the emulator on two SPS models\footnote{Implemented with the SPS code \textsc{fsps} \citep{conroy2009, conroy2010} with python bindings \textsc{python-fsps} \citep{foreman2014}.}: one relatively simple eight-parameter model targeting upcoming DESI observations (for which we emulate spectra), and the more flexible 14-parameter Prospector-$\alpha$ model from the recent \citet{leja2019} analysis (for which we emulate both spectra and photometry). For both models, we show that the emulator is able to deliver percent-level accuracy over broad parameter prior and wavelength ranges, and gives a factor $\sim 10^3-10^4$ speed-up over direct SPS model evaluation. Use of gradient-based inference methods enabled by the emulators will provide further reductions in the cost of inference under SPS models.

The structure of this paper is as follows: In \S \ref{sec:emulation} we outline the emulation framework. In \S \ref{sec:fomo}--\ref{sec:prospector} we validate the spectrum emulator on two SPS model parameterizations. In \S \ref{sec:photometry} we validate the photometry emulator for the Prospector-$\alpha$ model. We discuss the implications for current and future studies in \S \ref{sec:discussion}.
\section{SPECULATOR: Emulating stellar population synthesis}
\label{sec:emulation}
\begin{figure*}
\centering
\includegraphics[width = 17.5cm]{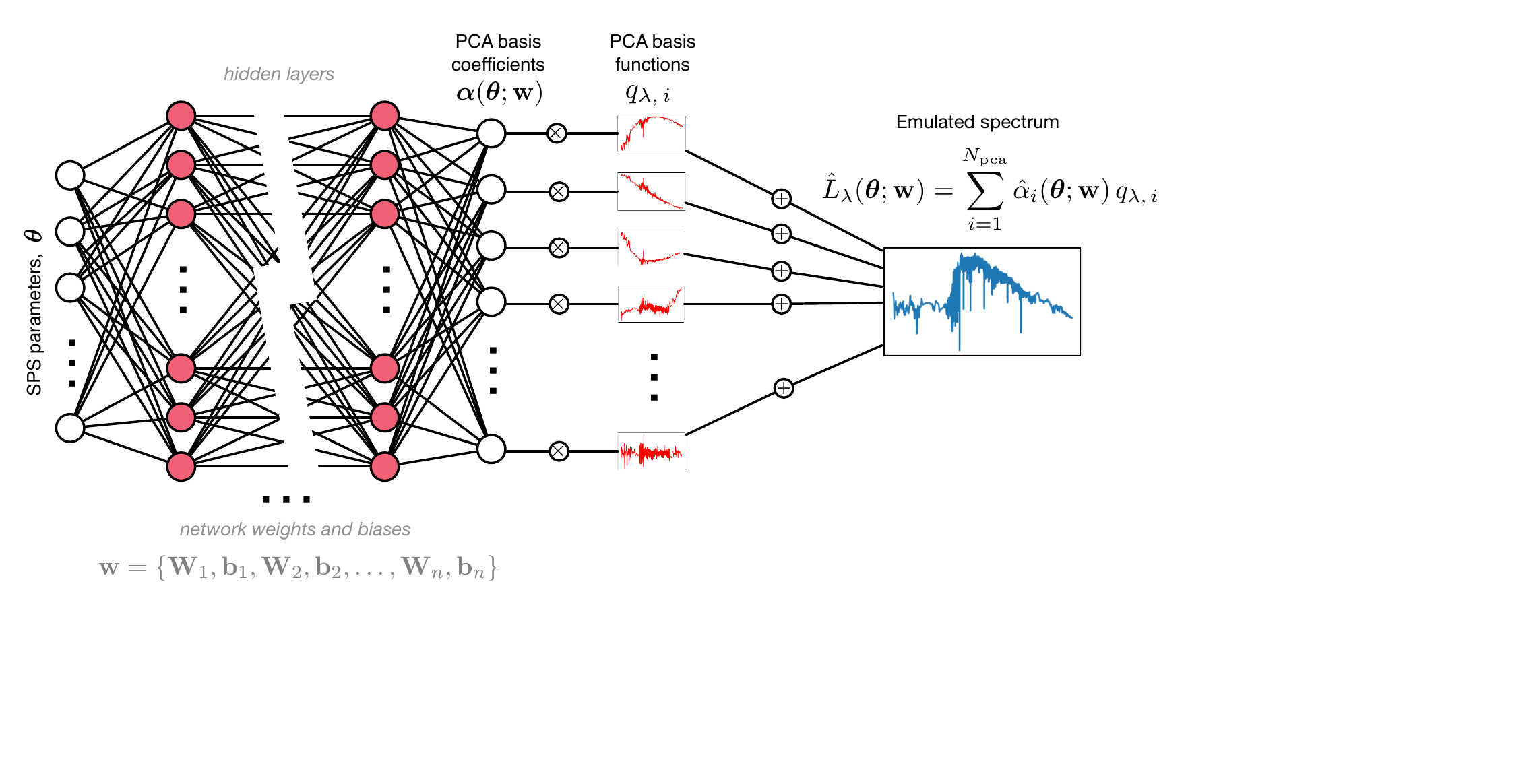}
\caption{Schematic of the PCA neural network emulator set-up. A dense neural network parameterizes the PCA basis coefficients as a function of the SPS model parameters (i.e., taking SPS parameters as input and predicting the basis coefficients). These basis coefficients are then multiplied by their respective PCA basis functions and summed to give the predicted spectrum.}
\label{fig:speculator}
\end{figure*}
In this section we describe the framework developed for fast emulation of SPS spectra (\S \ref{sec:emu}) and photometry (\S \ref{sec:emu_phot}). Some background knowledge of PCA and neural networks is assumed in this section; see e.g., \citet{bishop2006} for a comprehensive and pedagogical review. For previous work on representing spectra as interpolations over PCA bases, see \cite{czekala2015,kalmbach2017}.
\subsection{Notation}
\label{sec:sps}
We will denote galaxy SEDs by $l(\lambda; \btheta)\equiv l_\lambda$ (luminosity per unit wavelength) and log SEDs by $L_\lambda \equiv\ln\,l_\lambda$, for wavelength $\lambda$ and SPS model parameters $\btheta$. Photometric fluxes, denoted by $f_b(\btheta)$, for a given band-pass $b$ with filter $W_b(\lambda)$ and SPS parameters $\btheta$, are given by
\begin{align}
    f_b(\btheta) = \frac{1}{g^\mathrm{AB}4\pi (1+z)d_L^2(z)}\int_0^\infty l(\lambda/(1+z); \btheta)\,W_b(\lambda)d\lambda,
\end{align}
where $g^\mathrm{AB}$ is the AB flux normalization, $d_L(z)$ the luminosity distance for redshift $z$, and the filter is assumed to be normalized to unity, $\int W_b(\lambda)d\lambda = 1$. The associated apparent magnitudes are denoted by $m_b(\btheta)$.

The goal of emulation is to find an efficient representation for the galaxy spectra $l_\lambda(\btheta)$ or photometry $\{m_b(\btheta)\}$ as a function of the SPS model parameters that is as fast as possible to evaluate, whilst maintaining accuracy.
\subsection{Emulation of galaxy spectra}
\label{sec:emu}
\subsubsection{Parameterization considerations}
There are a couple of simplifications to the SED-emulation problem set-up that will make emulation significantly easier.

We will emulate the rest-frame SEDs only, redshifting (analytically) afterwards as needed. This is motivated by the fact that emulator is contingent on finding a compact PCA basis for galaxy SEDs; constructing this basis is greatly simplified when working with in the rest-frame only, i.e., without requiring that the basis can capture arbitrary stretches in wavelength. Meanwhile, emulating rest-frame SEDs only does not reduce functionality, since redshifted spectra can be obtained straightforwardly (and exactly) from the rest-frame SEDs.

Redshifting involves three transformations on the emulated rest-frame SEDs: a stretch by $\lambda\rightarrow\lambda/(1+z)$, re-scaling by $[(1+z)d_L(z)^2]^{-1}$, and adjusting the age of the Universe at the lookback time for a given redshift, $t_\mathrm{age}(z)$, so that the age of the stellar population is consistent with that lookback time. Therefore, $t_\mathrm{age}(z)$ must be included in the list of SPS parameters $\btheta$.

Similarly, we fix the total stellar-mass, $M$, for the emulated spectra to $1\,\mathrm{M}_\odot$ and scale the mass analytically afterwards as required (the total stellar-mass formed $M$ enters as a simple normalization of the SED). Hence, a galaxy spectrum for a given redshift $z$, total stellar-mass formed $M$, and SPS model parameters $\btheta$ can be obtained from the corresponding emulated rest-frame unit stellar-mass SED $l(\lambda ; \btheta)$ as
\begin{align}
\label{redshift}
    l(\lambda;\btheta, M, z) \rightarrow  l(\lambda/(1+z);\btheta)\vert_{t_\mathrm{age}(z)}&\, \frac{1}{(1+z)d_\mathrm{L}(z)^2}\, M.
\end{align}
\subsubsection{PCA neural network emulator framework}
A schematic overview of the PCA network emulator framework described below is given in Figure \ref{fig:speculator}, for reference throughout this section.

To build an emulator for a given SPS model parameterization, we begin by generating a training set of $N_\mathrm{train}$ galaxy SEDs $\{(L_\lambda, \btheta)_1, (L_\lambda, \btheta)_2, \dots, (L_\lambda, \btheta)_{N_\mathrm{train}}\}$ under the target SPS model, by drawing SPS parameters from the prior and computing the associated SEDs. 

From this training set, we construct a basis $\{q_{\lambda,\,i}\}$ for the SEDs by performing a PCA decomposition of the training spectra, and taking the first $N_\mathrm{pca}$ principal components as basis vectors. The number of PCA components retained is chosen such that the resulting PCA basis is comfortably able to recover the model SEDs at the desired accuracy (i.e., $\ll 1\%$ if we want to ensure that the errors associated with the PCA basis are a small fraction of the total error budget).

With the PCA basis $\{q_{\lambda,\,i}\}$ in hand, we model the (log) SED as a linear combination of the PCA basis functions,
\begin{align}
    L_\lambda(\btheta) = \sum_{i=1}^{N_\mathrm{pca}}\alpha_i(\btheta)\,q_{\lambda,\,i},
\end{align}
where the vector of coefficients $\boldsymbol{\alpha}(\btheta)$ are some unknown (non-linear) functions of the SPS parameters $\btheta$. The remaining step, then, is to learn some convenient parametric model $\hat{\boldsymbol\alpha}(\btheta;\vect{w})$ (with parameters $\vect{w})$ for the basis coefficients $\boldsymbol{\alpha}(\btheta)$ as a function of the SPS parameters. 

We parameterize the basis coefficients as a function of the model parameters by a dense fully-connected neural network with $n$ hidden layers, with $\{h_1, h_2,\dots,h_n\}$ hidden units and non-linear activation functions $\{a_1, a_2,\dots,a_n\}$ respectively, i.e., 
\begin{align}
\label{nn}
    \hat{\boldsymbol\alpha}(\btheta; \vect{w}) &= a_n(\mathbf{W}_n\vect{y}_{n-1} + \vect{b}_n), \nonumber \\
    \vect{y}_{n-1} &= a_{n-1}(\mathbf{W}_{n-1}\vect{y}_{n-2} + \vect{b}_{n-1}) \nonumber \\
    &\;\vdots \nonumber \\
    \vect{y}_1 &= a_1(\mathbf{W}_1\boldsymbol\theta + \vect{b}_1).
\end{align}
The weight matrices and bias vectors for each network layer are denoted by $\mathbf{W}_k\in\mathbb{R}^{h_k\times h_{k-1}}$ and $\vect{b}_k\in\mathbb{R}^k$, we use $\vect{w} = \{\mathbf{W}_k,\vect{b}_k\}$ as shorthand for the full set of weights and biases of the whole network, and $\mathbf{y}_k$ denotes the output from layer $k$.

Finally, to train the emulator we optimize the network parameters $\vect{w}$ by minimizing the loss function,
\begin{align}
    -\ln\,U(\vect{w} ; \{\btheta, \boldsymbol\alpha\}) = \frac{1}{N_\mathrm{train}} \sum_{m=1}^{N_\mathrm{train}} | \boldsymbol\alpha_m - \hat{\boldsymbol\alpha}(\btheta_m; \vect{w}) |^2,
\end{align}
where $\{\boldsymbol\alpha_m\}$ are the PCA basis coefficients for the SEDs $\{L_\lambda\}$ in the training set, and $\btheta_m$ the corresponding SPS model parameters for those training set members. This loss function is just the mean square error between neural network predicted and true PCA basis coefficients over the training set.

The emulator model is succinctly summarized by
\begin{align}
\label{emulator}
    \hat{\mathbf{L}}(\btheta) = \mathbf{Q}\,\hat{\boldsymbol\alpha}(\btheta ; \vect{w}),
\end{align}
where $\hat{\mathbf{L}}(\btheta) = (\hat{L}_{\lambda,1}(\btheta), \hat{L}_{\lambda,2}(\btheta), \dots, \hat{L}_{\lambda,{N_\lambda}}(\btheta))$ is the emulated SED for parameters $\btheta$, $Q_{\lambda i} = q_{\lambda,i}$ is the set of basis functions, and $\hat{\boldsymbol\alpha}(\btheta ; \vect{w})$ is given by Eq. \eqref{nn}. The neural network emulator is specified entirely by the set of matrices and non-linear activation functions $\{\mathbf{W}_k, \vect{b}_k, \mathbf{Q}, a_k\}$. Calculating an emulated SPS model spectrum using Eqs. \eqref{emulator} and \eqref{nn} is hence reduced to a series of linear matrix operations, and passes through simple non-linear (e.g., tanh) activation functions. Furthermore, the neural network in Eq. \eqref{nn} is straightforwardly differentiable (by the chain rule), so derivatives of the model spectra with respect to the SPS parameters are readily available. We highlight that implementation of the trained emulator using Eqs. \eqref{nn} and \eqref{emulator} is simple, so incorporating the trained emulator into existing (or future) analysis codes should be straightforward.

In the limit of a large PCA basis, large training set, and complex neural network architecture, the emulator described above can represent any (deterministic) SPS model to arbitrary precision. However, the power of this emulation framework comes from the fact that -- as we will demonstrate in the following sections -- a relatively small PCA basis and neural network architecture can achieve percent-level precision over broad parameter ranges, even for relatively complex SPS parameterizations. It is this fact that allows the emulator to achieve such significant speed ups.
\subsubsection{Discussion}
The use of neural networks in this context is solely as a convenient parametric model for an unknown function that we want to learn, in a situation where the dimensionality is too high to make direct interpolation efficient. Neural networks have a number of useful features that make them well-suited to this sort of emulation task. The universal approximation theorem tells us that a neural network with a single hidden layer and finite number of nodes can approximate any continuous function on compact subsets of $\mathbb{R}^n$ under some mild assumptions about the activation function \citep{csaji2001}. Their derivatives can be computed efficiently (by backpropagation), making for efficient training. Once trained, they are straightforward and fast to evaluate, and importantly the computational cost of evaluation is fixed ahead of time and independent of the size of the training set (in contrast to Gaussian processes\footnote{For use of PCA and Gaussian processes in a similar context, see \citet{czekala2015}.}, where the cost of evaluation na\"{i}vely scales as $N^3$ with the training set size).

In this study we show that relatively simple dense fully-connected network architectures are able to perform well in the context of SPS emulation. However, for more complex SPS models than those considered here, or where fidelity requirements are very high, more sophisticated architectures may prove beneficial (for more discussion see \S \ref{sec:discussion}).

We note that training an emulator on a given SPS parameterization is performed over some pre-determined prior ranges for the parameters. Care should be taken to train the emulator over well-chosen priors in the first instance, since emulated SEDs outside of the pre-determined prior ranges of the training set should not be expected to be reliable.
\subsection{Emulation of galaxy photometry}
\label{sec:emu_phot}
\begin{figure}
\centering
\includegraphics[width = 7.5cm]{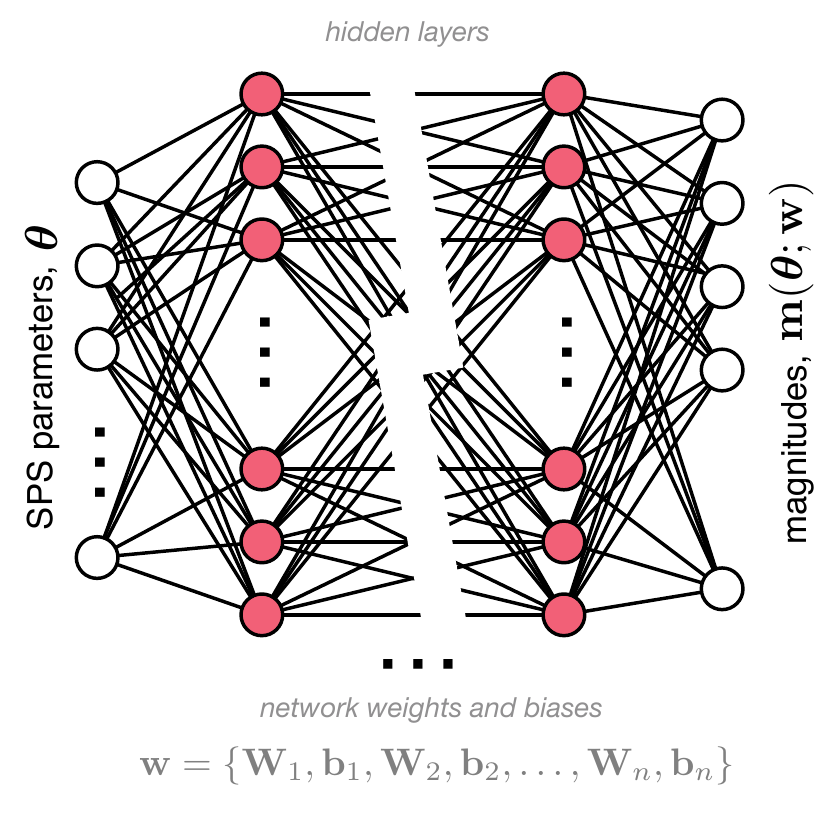}
\caption{Schematic of the emulator set-up for photometry under SPS models; the magnitudes (for some chosen set of band-passes) as a function of the SPS model parameters are parameterized as a dense fully-connected neural network (c.f., Eq. \eqref{nn2}).}
\label{fig:photulator}
\end{figure}
For applications where photometry rather than spectra are the primary target, it makes sense to emulate the photometry directly, i.e., learn a compact model for the fluxes or magnitudes for some set of filters, as a function of the SPS parameters. Emulating photometry presents a simpler problem than emulating spectra: the number of bands of interest is typically $\mathcal{O}(10)$ (or fewer), so no basis construction or dimensionality reduction is necessary.

To emulate photometry for some set of band-passes $\{b_1, b_2,\dots, b_k\}$ under a given SPS model, we parameterize the magnitudes $\mathbf{m}(\btheta) = (m_{b_1}(\btheta), m_{b_2}(\btheta),\dots,m_{b_k}(\btheta))$ by a dense fully-connected neural network, i.e. (Figure \ref{fig:photulator}),
\begin{align}
\label{nn2}
    \hat{\mathbf{m}}(\btheta; \vect{w}) &= a_n(\mathbf{W}_n\vect{y}_{n-1} + \vect{b}_n), \nonumber \\
    \vect{y}_{n-1} &= a_{n-1}(\mathbf{W}_{n-1}\vect{y}_{n-2} + \vect{b}_{n-1}) \nonumber \\
    &\;\vdots \nonumber \\
    \vect{y}_1 &= a_1(\mathbf{W}_1\boldsymbol\theta + \vect{b}_1),
\end{align}
where $\hat{\mathbf{m}}(\btheta; \vect{w})$ denotes the neural network emulated photometry. As before, the weight matrices and bias vectors for each network layer are denoted by $\mathbf{W}_k\in\mathbb{R}^{h_k\times h_{k-1}}$ and $\vect{b}_k\in\mathbb{R}^k$, we use $\vect{w} = \{\mathbf{W}_k,\vect{b}_k\}$ as shorthand for the full set of weights and biases of the whole network.
\subsection{Activation function choice for neural SPS emulation}
\label{sec:activation}
We find that SPS spectra and photometry as functions of the model parameters are mostly smooth, but exhibit some non-smooth features. In particular, the behavior as a function of stellar and gas metallicity parameters exhibits discontinuous changes in gradient. When considering neural network architecture choices for SPS emulation, it is therefore advantageous to choose activation functions that are able to capture both smooth features and sharp gradient changes; well-chosen activation functions will allow us to achieve higher fidelity emulation with smaller (faster) network architectures.

To this end, we adopt activation functions of the following form,
\begin{align}
\label{activation}
    a(\mathbf{x}) = \left[\boldsymbol{\gamma} + (1+e^{-\boldsymbol\beta\odot\mathbf{x}})^{-1}(1-\boldsymbol{\gamma})\right]\odot\mathbf{x},
\end{align}
where $\boldsymbol{\gamma}$ and $\boldsymbol\beta$ are included as additional free parameters of the network to be trained alongside the network weights and biases. This activation function is able to cover smooth features (small $\beta$), and sharp changes in gradient (as $\beta\rightarrow\infty$). In experiments, we find that activation funcions of this form outperform other popular neural network activation choices for the SPS emulation problem (including tanh, sigmoid, ReLU and leaky-ReLU; see \citealp{Nwankpa2018} for recent trends in activation function choice). Non-linear activation functions of the form Eq. \eqref{activation} are hence adopted throughout this work.
\subsection{Target accuracy for SPS emulation}
Whilst a great deal of progress has been made in reducing modeling uncertainties associated with stellar population synthesis, some fundamental uncertainties remain (e.g., the effect of binaries and rotation on the ionizing photon production from massive stars \citealp{choi2017}; for a review of SPS model uncertainties see \citealp{conroy2013}). When analyzing galaxies under SPS models it is therefore common practice to assume an error floor of $2$--$5\%$ on the SEDs or photometry, to account for the theoretical SPS model uncertainties (e.g., \citealp{leja2019}). On the observational side, for photometry it is also common practice to put an error floor (typically 5\%) on the measured fluxes to account for systematic uncertainties in the photometric calibration (e.g., \citealp{Muzzin2013a, Chevallard2016, Pacifici2016, Belli2019, Carnall2019}).

This context provides a natural accuracy target for SPS emulation (for both spectra and photometry): $\lesssim 5\%$ accuracy, or, $\ll 5\%$ if we want to ensure the emulator error is a small fraction of the total error budget. Whilst this covers a range of use cases, we note that for analysis of high S/N spectra under very complex SPS models, the fidelity requirements may be more like $\ll 1\%$ (see \S \ref{sec:discussion} for discussion).
\section{Validation I: DESI model spectra}
\label{sec:fomo}
In this section, we demonstrate and validate the emulator on a relatively simple eight-parameter SPS parameterization. The model is outlined in \S \ref{sec:fomo_model}, the emulator set-up described in \S \ref{sec:fomo_emulator}, and validation tests and performance discussed in \S \ref{sec:fomo_validation}-\ref{sec:fomo_performance}.
\subsection{Model and priors}
\label{sec:fomo_model}
\begin{figure*}
\centering
\includegraphics[width = 17.5cm]{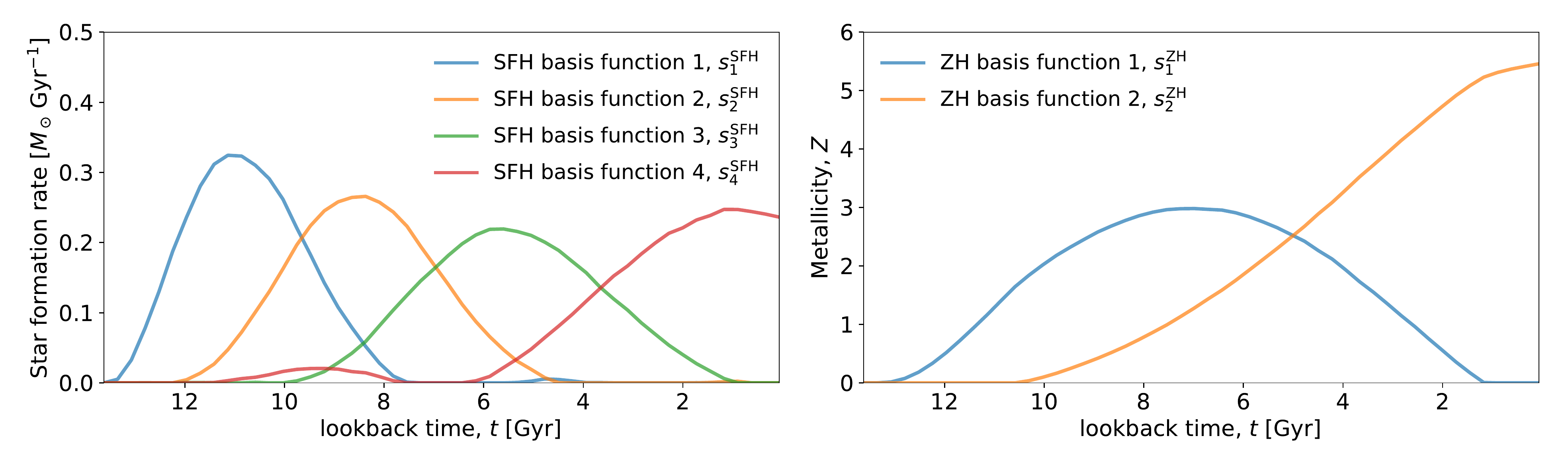}
\caption{Basis functions for the star formation history (left) and metallicity history (right) for the DESI model (see \S \ref{sec:fomo_model}). The SFH basis functions are normalized such that the total mass formed is one solar mass. The metallicity components are unnormalized, but the values refer to the mass fraction in metals ($Z_\odot$ = 0.019).}
\label{fig:fomo_sfh_zh}
\end{figure*}
Our first model (hereafter, the DESI model) is motivated by upcoming analyses of large numbers of optical, low signal-to-noise (S/N) spectra being collected by current and future surveys. The specifics of the model presented in this section are targeted at the analysis of low-redshift spectra for the upcoming DESI Bright Galaxy Survey \citep[BGS;][]{aghamousa2016}. The BGS will be a flux-limited 
survey that will target ${\gtrsim}10$ million galaxies with
$z \lesssim 0.45$ over $14,000~{\rm deg}^2$. It will 
measure spectra over a wavelength range between $360$ to
$980\mathrm{nm}$ with a resolution $R = \lambda / \Delta \lambda$ between 2000 and 5500, depending on the wavelength. Individual spectra will have a median S/N of $\sim2-3$ 
per pixel. The key features and free parameters of the model, and associated prior ranges, are as follows.

We model the star-formation and chemical enrichment histories as a function of lookback time as linear combinations of a set of pre-computed basis functions (Figure \ref{fig:fomo_sfh_zh}). The shape and number of basis functions were determined by applying a non-negative matrix factorization to the star-formation and chemical enrichment histories of galaxies above $10^9$ M$_\odot$ in the Illustris simulation \citep{Vogelsberger2014}. We sought to construct a basis with the minimal number of components that would reconstruct the history of galaxies, and therefore their optical spectra, to an accuracy dictated by the typical DESI S/N. In practice, the chosen basis has a dependence on the optical colours of the galaxies. The basis used here is an indicative example of what will be used to analyse DESI spectra; further details are given in Tojeiro et al. (in prep).

The star formation history\footnote{i.e., stellar mass formed per unit time, $[\mathrm{M}_\odot \mathrm{Gyr^{-1}}]$.} for a galaxy at redshift $z$ is implemented as a linear combination of four SFH basis functions $\{s^\mathrm{SFH}_i(t)\}$ (shown in Figure \ref{fig:fomo_sfh_zh})
\begin{align}
    \mathrm{SFH}(t; t_\mathrm{age}(z)) = \sum_{i=1}^4\beta^\mathrm{SFH}_i\,\frac{s^\mathrm{SFH}_i(t)}{\int_0^{t_\mathrm{age}(z)}s^\mathrm{SFH}_i(t)dt},
\end{align}
where the SFH basis coefficients $\{\beta^\mathrm{SFH}_i\}$ are free parameters of the model, the basis functions are normalized to unity over the age of the Universe at the lookback time of the galaxy $t_\mathrm{age}(z)$, and time runs from $0$ to $t_\mathrm{age}(z)$. We train the emulator over a flat-Dirichlet prior for the basis coefficients, i.e., a uniform prior over all combinations of basis coefficients under the constraint that $\sum_{i=1}^4\beta^\mathrm{SFH}_i=1$ (ensuring that the total SFH is normalized to unity for the emulated spectra).

The metallicity enrichment history (ZH) is similarly parameterized as a linear combination of two basis functions $\{s^\mathrm{SFH}_i(t)\}$ (shown in Figure \ref{fig:fomo_sfh_zh})
\begin{align}
    \mathrm{ZH}(t) = \sum_{i=1}^2\gamma^\mathrm{ZH}_i\,s^\mathrm{ZH}_i(t),
\end{align}
where again the ZH basis coefficients $\{\gamma^\mathrm{ZH}_i\}$ are free parameters of the model, and time runs from $0$ to $t_\mathrm{age}(z)$. We take uniform priors for the ZH basis coefficients, $\gamma^\mathrm{ZH}_i\in[6.9\times 10^{-5}, 7.33\times 10^{-3}]$.

Dust attenuation is modelled using the \citet{calzetti2000} attenuation curve, with the optical depth $\tau_\mathrm{ISM}$ as a free parameter with a uniform prior $\tau_\mathrm{ISM}\in[0,3]$.

The eight model parameters, their physical meanings, and associated priors are summarized in Table \ref{tab:fomo}.
\begin{table*}
\centering
\scalebox{0.95}{
\begin{tabularx}{\textwidth}{ccc}
\toprule
Parameter & Description & Prior \tabularnewline
\hline
$\beta^\mathrm{SFH}_1,\,\beta^\mathrm{SFH}_2,\,\beta^\mathrm{SFH}_3,\,\beta^\mathrm{SFH}_4$ & Star formation history basis function coefficients & flat-Dirichlet\tabularnewline
$\gamma^\mathrm{ZH}_1,\,\gamma^\mathrm{ZH}_2$ & Metallicity enrichment history basis function coefficients & Uniform $[6.9\times 10^{-5}, 7.3\times 10^{-3}]$\tabularnewline
$t_\mathrm{age}$ & Age of Universe at lookback-time of the galaxy & Uniform $[9.5, 13.7]\,\mathrm{Gyr}$\tabularnewline
&&(equivalent to $0 < z < 0.4$)\tabularnewline
$\tau_\mathrm{ISM}$ & Dust optical depth (\citealp{calzetti2000} attenuation model) & Uniform $[0, 3]$\tabularnewline
\hline
\end{tabularx}}
\caption{Summary of SPS model parameters and their respective priors for the DESI model (\S \ref{sec:fomo_model}).}
\label{tab:fomo}
\end{table*}
\subsection{Emulation}
\label{sec:fomo_emulator}
We generated a training and validation set of $5\times 10^5$ and $10^5$ SEDs respectively, for model parameters drawn from their respective priors (see Table \ref{tab:fomo}) and covering the wavelength range $200$ to $1000\,\mathrm{nm}$.% The model was implemented using the SPS code \textsc{fsps}.

 The PCA basis was constructed by performing a PCA decomposition of all of the training SEDs\footnote{Performing a PCA decomposition over large training sets can be memory intensive. Here we used \textsc{scikit-learn}'s ``incremental PCA", which constructs a PCA basis while only processing a few training samples at a time, keeping the memory requirements under control.}. We choose the number of PCA components to keep in the basis such that the basis is able to describe the validation SEDs to $\ll 1\%$ accuracy over the whole wavelength range and parameter volume. Figure \ref{fig:fomo_pca_variance} shows the fractional error distribution of the validation spectra represented in the PCA basis with $20$ components retained; the $20$ component basis is able to describe the SEDs to $\lesssim 0.5\%$ accuracy over the whole wavelength and parameter prior range. Note that the PCA basis is constructed for log SEDs, but accuracy in Figure \ref{fig:fomo_pca_variance} is displayed in linear space.

The PCA basis coefficients are parameterized by a dense neural network with two hidden layers of $256$ hidden units, with non-linear activation functions (Eq. \eqref{activation}) on all expect the output layer, which has linear activation. The network is implemented in \textsc{tensorflow} \citep{tensorflow2015-whitepaper} and trained with the stochastic gradient descent optimizer \textsc{adam} \citep{kingma2014adam}. Overfitting is mitigated by early-stopping\footnote{The training set is split $9:1$ into training and validation subsets, the networks are trained by minimizing the loss for the training subset only, but the loss for the validation subset is tracked during training. Overfitting is observed when the validation loss stops improving, whilst the training loss continues to decrease. Training is terminated when the loss of the validation set ceases to improve over $20$ training epochs.}.

Network training is performed on a Tesla K80 GPU\footnote{Freely available with Google Colab \url{https://colab.research.google.com/}.} and takes of the order of a few minutes for the network architecture and training set described above; the computational cost of building the emulator is overwhelmingly dominated by performing the direct SPS computations (using \textsc{fsps}) to generate the training set ($\sim$10 hours compared to minutes).
\subsection{Results and validation}
\label{sec:fomo_validation}
\begin{figure*}
\centering
\includegraphics[width = 17.5cm]{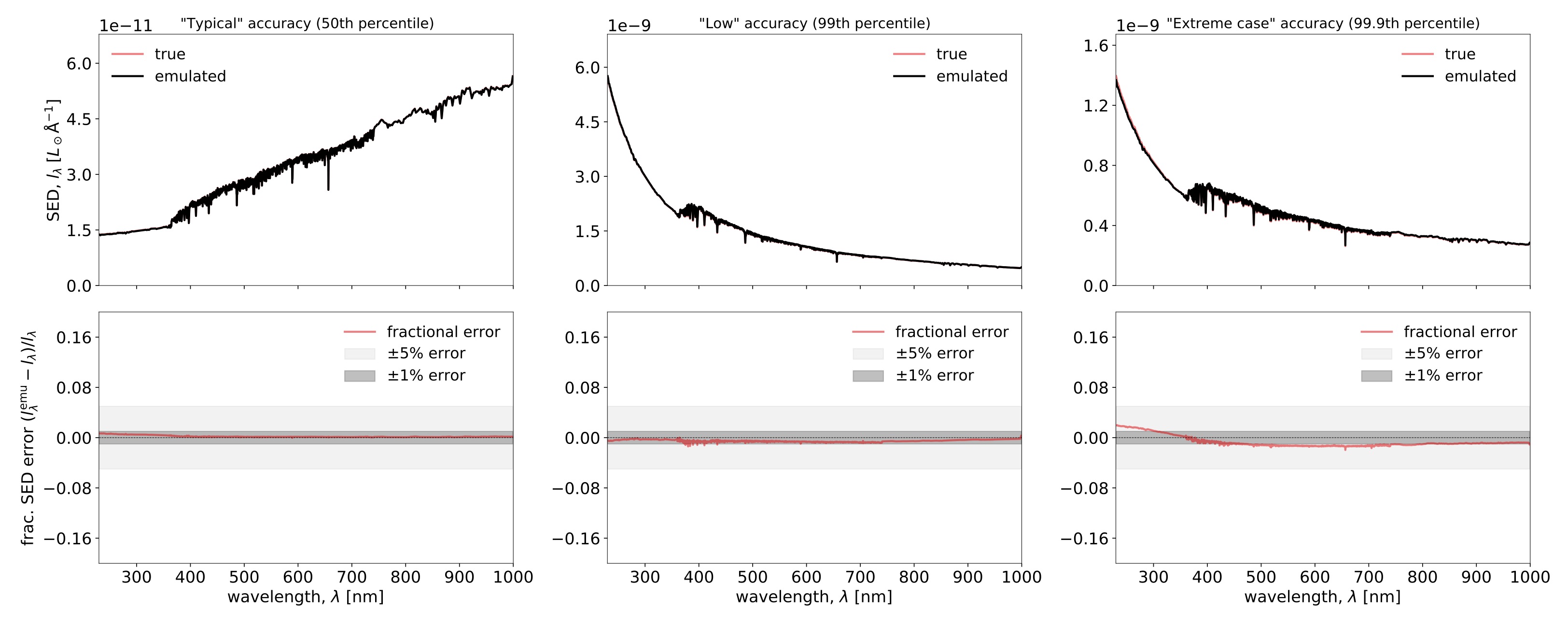}
\includegraphics[width = 17cm]{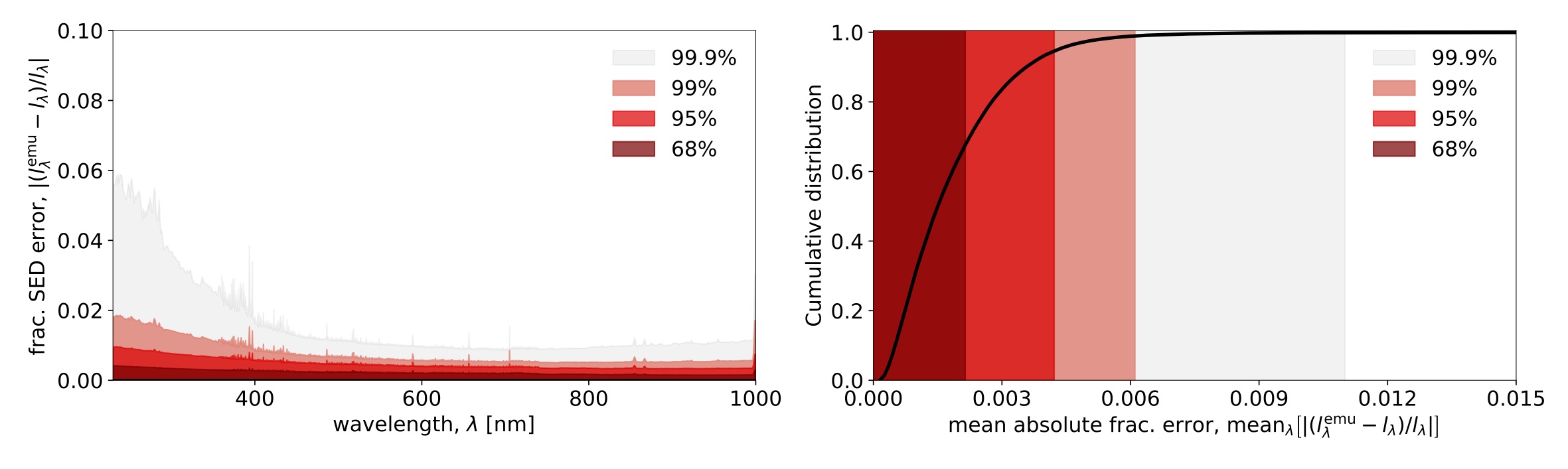}
\caption{Validation of the emulator for the DESI model (\S \ref{sec:fomo}). Top figure: ``typical", ``low" and ``extreme case" accuracy of the emulated SEDs from a validation set of $10^5$ spectra generated with parameters drawn from the prior (Table \ref{tab:fomo}). These cases correspond to the 50th, 99th and 99.9th percentiles of the mean (absolute) fractional error between emulated and true SED (over the wavelength range). Bottom left: 68 (dark red), 95 (red), 99 (salmon) and 99.9 (grey) percentiles of the fractional emulator error as a function of wavelength. Bottom right: cumulative distribution (blue) and 68 (dark red), 95 (red), 99 (salmon) and 99.9 (grey) percentiles of the mean (absolute) fractional errors (over the wavelength range). We see that the emulator is broadly accurate to $\lesssim 1\%$, with a small fraction (less than one percent) of validation samples having errors at the few-percent level or more at the lower end of the wavelength range.}
\label{fig:fomo_sed_accuracy}
\end{figure*}

For validating the trained emulator, we generated $10^5$ SEDs for model parameters drawn from the prior, and compared the emulated and exact SPS spectra for this validation set. The results are summarized in Figure \ref{fig:fomo_sed_accuracy}. The upper panels show typical, low and extreme case performance of the emulator, taken as the $50$th, $99$th, and $99.9$th percentiles of the mean (absolute) fractional error per SED (over the full wavelength range). The bottom left panel shows the $68$, $95$, $99$ and $99.9$ percent intervals of the fractional error as a function of wavelength, and the bottom right panel shows the cumulative distribution of the mean (absolute) fractional error for the validation samples (over the wavelength range). Note that the emulator is trained on the PCA coefficients of log SEDs, but accuracy is shown in Figure \ref{fig:fomo_sed_accuracy} in linear space.

The emulator is accurate at the $<1\%$ level over the full wavelength range for $>99\%$ of the SEDs in the validation set. A small fraction (less than one percent) of validation samples have errors at the few-percent level at the shortest wavelengths. We note that this small number of ``outliers" occur where the recent star formation history turns on/off and the SEDs are very sensitive to the most-recent SFH coefficients. Whilst even in these cases the emulator errors are acceptable, they may be further improved by re-parameterization of the SFH, or better sampling of the prior volume in this part of parameter space.

\begin{figure}
%\centering
\includegraphics[width = 8.5cm]{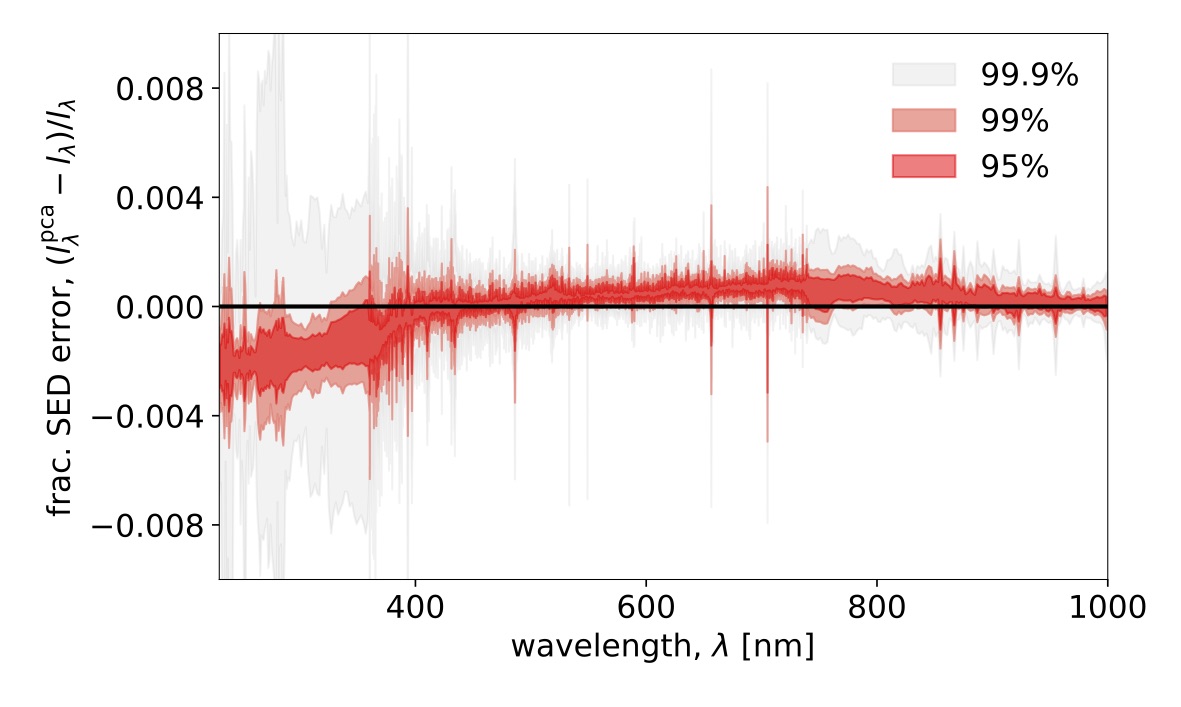}
\caption{Validation of the PCA basis for the DESI model (\S \ref{sec:fomo}). Shown are the central 95\% (red), 99\% (salmon) and 99.9\% (grey) intervals of the fractional errors on the DESI model spectra represented in the basis of the first $20$ PCA components. The $20$ PCA component basis is able to describe the model spectra to $\lesssim 0.5\%$ accuracy over the whole wavelength range and parameter volume.}
\label{fig:fomo_pca_variance}
\end{figure}

There are two distinct sources of emulator error: the adequacy of the PCA basis, and the accuracy of the neural network in learning the PCA basis coefficients as functions of the SPS parameters. Comparing Figures \ref{fig:fomo_pca_variance} and \ref{fig:fomo_sed_accuracy} (bottom left), we see that the error budget in this case is dominated by the neural network rather than the PCA basis. Accuracy could hence be further improved with a larger neural network architecture (accompanied by a larger training set if necessary), at the cost of some reduction in the performance gain (since a larger network will be more expensive to evaluate).
\subsection{Computational performance}
\label{sec:fomo_performance}
With the network architecture described above (\S \ref{sec:fomo_emulator}), we find that the trained emulator is able to generate predicted SEDs a factor of $10^4$ faster than direct SPS computation with \textsc{fsps} on the same (CPU) architecture. 

Implementation in \textsc{tensorflow} allows the emulator to automatically be called from a GPU, allowing for easy exploitation of GPU-enabled parallelization. Generating $10^6$ emulated SEDs takes around $\sim2\,\mathrm{s}$ on a Tesla K80 GPU, compared to $\sim 0.2\,\mathrm{s}$ per direct SPS computation on an Intel i7 CPU; an overall effective factor of $10^5$ speed-up.

When inferring SPS model parameters from galaxy observations, additional performance gains are expected from the use of gradient-based inference and optimization methods that are enabled by the emulator (which has readily available derivatives). We leave investigation of these extra gains to future work.
\section{Validation II: Prospector-$\alpha$ spectra}
\label{sec:prospector}
In this section we demonstrate and validate the spectrum emulator on a more flexible 14-parameter SPS parameterization -- the Prospector-$\alpha$ model \citep{leja2017, leja2018, leja2019}. The model is outlined in \S \ref{sec:prospector_model}, the emulator set-up described in \S \ref{sec:prospector_emulator}, and validation tests and results discussed in \S \ref{sec:prospector_validation}-\ref{sec:prospector_performance}.
\subsection{Model and priors}
\label{sec:prospector_model}
The Prospector-$\alpha$ model includes a non-parametric star formation history, a two-component dust attenuation model with a flexible attenuation curve, variable stellar and gas-phase metallicity, dust emission powered via energy balance, and emission from a dusty AGN torus. Nebular line and continuum emission is generated using CLOUDY \citep{ferland2013} model grids from \citet{byler2017}. MIST stellar evolution tracks and isochrones are assumed \citep{choi2016, dotter2016}, based on MESA \citep{paxton2010, paxton2013, paxton2015}.

The model has been tested and calibrated on local galaxies \citep{leja2017, leja2018}, and recently used to analyze a sample of $\sim 60,000$ galaxies from the 3D-HST photometric catalog \citep{skelton2014} over $0.5<z<2.5$ \citep{leja2019}. The model is described in detail in \citet{leja2017, leja2018, leja2019}; we review the salient features, model parameters and associated priors below. A summary of model parameters and priors is given in Table \ref{tab:prospector}.

The star formation history is modelled as piece-wise constant, with seven time bins spaced as follows. Two bins are fixed at $[0,30]$ $\mathrm{Myr}$ and $[30,100]$ $\mathrm{Myr}$ to capture recent SFH. A third bin is placed at the other end at $[0.85, 1]\,t_\mathrm{age}$, where $t_\mathrm{age}$ is the age of the Universe at the lookback time of the galaxy, to model the oldest star formation. The remaining four bins are spaced equally in logarithmic time between $100$ $\mathrm{Myr}$ and $0.85\,t_\mathrm{age}$. The six ratios of the logarithmic star formation rate (SFR) in adjacent SFH bins $\{r_\mathrm{SFH}^i\}$ are included as free model parameters. Following \citet{leja2017, leja2018, leja2019} we take independent Student's-$t$ priors on the log SFR ratios (see Table \ref{tab:prospector}). This prior is chosen to allow similar transitions in the SFR as seen in the Illustris hydrodynamical simulations \citep{vogelsberger2014a, vogelsberger2014b, torrey2014, diemer2017}, although care is taken to ensure a wider range of models is allowed than is seen in those simulations.

A single stellar metallicity is assumed for all stars in a galaxy. The observed stellar mass-stellar metallicity relationship from $z = 0$ Sloan Digital Sky Survey (SDSS) data \citep{gallazzi2005} is used to motivate the metallicity prior. For a given stellar-mass, the stellar-metallicity prior is taken to be a truncated normal with limits\footnote{Set by the range of the MIST stellar evolution tracks.} $−1.98 < \mathrm{log}(Z/Z\odot) < 0.19$, mean set to the \citet{gallazzi2005} $z = 0$ relationship, and standard deviation taken to be twice the observed scatter about the $z=0$ relationship (to allow for potential redshift evolution in the mass-metallicity relation).

As discussed in \S \ref{sec:emulation} we fix the integral normalization of the SFH to $1\,\mathrm{M}_\odot$ for the spectra in the training set, and stellar-mass can then be set by adjusting the normalization of the emulated spectra. However, because in this case the metallicity prior is taken to be mass-dependent, we sample total stellar-mass formed from a log uniform prior from $10^7\mathrm{M}_\odot$ to $10^{12.5}\mathrm{M}_\odot$ first (for the purpose of sampling from the metalliticy prior correctly), and then renormalize the spectra to $1\,\mathrm{M}_\odot$ afterwards when training the emulator.

Gas-phase metallicity is decoupled from the stellar metallicity and allowed to vary (uniformly) between $−2 < \mathrm{log}(Z_\mathrm{gas}/Z_\odot) < 0.5$.

Dust is modelled with two components -- birth cloud and diffuse dust screens -- following \citet{charlot2000} (see \citealp{leja2017} for details). The birth cloud ($\tau_1$) and diffuse ($\tau_2$) optical depths are free model parameters, with truncated normal priors: $\tau_2\sim \mathcal{N}(0.3, 1)$ with limits $\tau_2\in[0,4]$, and $\tau_1 / \tau_2 \sim\mathcal{N}(1, 0.3)$ with limits $\tau_1/\tau_2\in[0,2]$. The power law index of the \citet{calzetti2000} attenuation curve for the diffuse component is also included as a free model parameter, with a uniform prior $n\in[-1, 0.4]$.

AGN activity is modelled as described in \citet{leja2018}, with the fraction of the bolometric luminosity from the AGN $f_\mathrm{AGN}$ and optical depth of the AGN torus $\tau_\mathrm{AGN}$ as free parameters with log-uniform priors $\ln\,f_\mathrm{AGN}\in[\ln(10^{-5}), \ln(3)]$  and $\ln\,\tau_\mathrm{AGN}\in[\ln(5), \ln(150)]$ respectively.

The model parameters, their physical meanings, and associated priors are summarized in Table \ref{tab:prospector}.
\begin{table*}
\centering
\scalebox{0.95}{
\begin{tabularx}{\textwidth}{ccc}
\toprule
Parameter & Description & Prior \tabularnewline
\hline
$M$ & Total stellar-mass formed & Log-Uniform $[10^7, 10^{12.5}]\mathrm{M}_\odot$\tabularnewline
${r_\mathrm{SFH}^1,\dots,r_\mathrm{SFH}^6}$ & Ratio of log-SFR between adjacent bins & Clipped Student's-$t$: $\sigma=0.3$, $\nu=2$, $|r_\mathrm{SFH}^i| \leq 5$\tabularnewline
$t_\mathrm{age}$ & Age of Universe at the lookback-time of galaxy & Uniform $[2.6, 13.7]\,\mathrm{Gyr}$, ($0 < z < 2.5$)\tabularnewline
$\tau_2$ & Diffuse dust optical depth & Normal $\mu = 0.3,\,\sigma=1$, min=0, max=4\tabularnewline
$\tau_1$ & Birth-cloud optical depth & Truncated normal in $\tau_1/\tau_2$\tabularnewline
&&$\mu = 1,\,\sigma=0.3$, min=0, max=2\tabularnewline
$n$ & Index of \citet{calzetti2000} dust attn. curve & Uniform $[-1, 0.4]$\tabularnewline
$\ln\,(Z_\mathrm{gas}/Z_\odot)$ & Gas phase metallicity & Uniform $[-2, 0.5]$\tabularnewline
$f_\mathrm{AGN}$ & Fraction of bolometric luminosity from AGN & Log-Uniform $[10^{-5}, 3]$\tabularnewline
$\tau_\mathrm{AGN}$ & Optical depth of AGN torus & Log-Uniform $[5, 150]$\tabularnewline
$\ln\,(Z/Z_\odot)$ & Stellar metallicity & Truncated normal with $\mu$ and $\sigma$ from \tabularnewline
&&\citet{gallazzi2005} mass-metallicity relation (see \S \ref{sec:prospector}),\tabularnewline
&&limits min=-1.98, max=0.19 \tabularnewline
$z$ & Redshift & Uniform [0.5, 2.5] \tabularnewline
\hline
\end{tabularx}}
\caption{Summary of SPS model parameters and their respective priors for the Prospector-$\alpha$ model (\S \ref{sec:prospector_model}). Note that for emulating spectra under this model (\S \ref{sec:prospector}), generated training spectra are computed in the rest-frame (but over a range of values for $t_\mathrm{age}$), and renormalized such that they correspond to $M=1\mathrm{M}_\odot$ (see \S \ref{sec:emulation} for motivation). When emulating photometry under this model (\S \ref{sec:photometry}), $M$ and $z$ are kept as free parameters to be emulated over.}
\label{tab:prospector}
\end{table*}
\subsection{Emulation}
\label{sec:prospector_emulator}
We generated a training and validation set of $2\times 10^6$ and $10^5$ SEDs respectively\footnote{We used a larger training set for the Prospector-$\alpha$ compared to the DESI model, owing to the larger parameter space. Training set sizes for both models were chosen so that they could be generated in $\lesssim$ days and achieved percent-level accuracy upon validation. For more discussion on optimization of training set sizes see \S \ref{sec:discussion}.}, for model parameters drawn from the prior (see Table \ref{tab:prospector}) and covering the wavelength range $100\,\mathrm{nm}$ to $30\,\mu\mathrm{m}$ (using the SPS code \textsc{fsps}). 

For emulating higher-dimensional SPS models over very broad wavelength ranges, such as this case, it is advantageous to split the emulation task into a number of wavelength sub-ranges, which can be stitched together afterwards. Here, we will emulate $100-400\,\mathrm{nm}$ (UV), $400-1100\,\mathrm{nm}$ (optical-NIR) and $1100\,\mathrm{nm}-30\,\mu\mathrm{m}$ (IR) separately. We find in experiments that without splitting into wavelength sub-ranges, more PCA components are required (in total) to achieve the same consistent accuracy across the full wavelength range. Furthermore, from the perspective of training the neural networks, emulating relatively smaller PCA bases (for each wavelength sub-range) represents an easier learning task compared to emulating a single large ($>100$ component) basis. This means that relatively smaller networks can be used for each sub-range, requiring less training data and being faster to evaluate once trained. We do not find any evidence for discontinuities in the emulated spectra at the boundaries between wavelength regions (within the accuracy of the emulator at the boundaries; Figure \ref{fig:prospector_sed_accuracy}). 

\begin{figure*}
\includegraphics[width = 18cm]{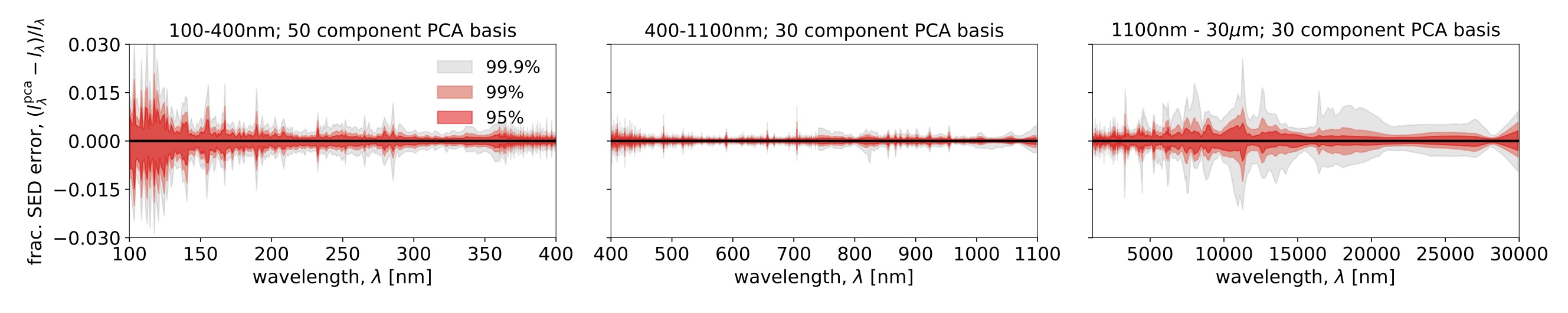}
\caption{Validation of the PCA basis for the Prospector-$\alpha$ model (\S \ref{sec:prospector}). Shown are the central 95 (red), 99\% (salmon) and 99.9\% (grey) intervals for the fractional errors on the $10^5$ validation spectra represented in the basis of the first 50, 30 and 30 PCA components for UV, optical-NIR and IR wavelength ranges respectively. The basis is able to capture the Prospector-$\alpha$ model spectra to $\lesssim 1\%$ accuracy over the entire wavelength and parameter ranges.}
\label{fig:prospector_pca_variance}
\end{figure*}

The PCA basis was constructed as before by performing a PCA decomposition of all of the training SEDs (for the three wavelength ranges separately), and the number of PCA components retained chosen such that the resulting basis is able to capture the (validation) SEDs with $\lesssim 1\%$ level accuracy. Figure \ref{fig:prospector_pca_variance} shows the distribution of errors on the validation SEDs for the PCA basis with $50$ components for UV, and $30$ components for optical-NIR and IR respectively. This basis is sufficient to describe the SEDs to $\lesssim 1\%$ over the full wavelength range and parameter volume. The errors can be reduced further by increasing the size of the PCA basis, but are sufficient for our current purposes. Note that the PCA basis was constructed for log SEDs, but accuracy shown in Figure \ref{fig:prospector_pca_variance} in linear space.

The basis coefficients for each wavelength range are parameterized by a dense neural network with three hidden layers of $256$ hidden units, with non-linear activation functions (Eq. \eqref{activation}) on all hidden layers, and linear activation on the output. Network implementation and training follows exactly as described in \S \ref{sec:fomo_emulator}.
\subsection{Results and validation}
\label{sec:prospector_validation}
\begin{figure*}
\centering
\includegraphics[width = 17.5cm]{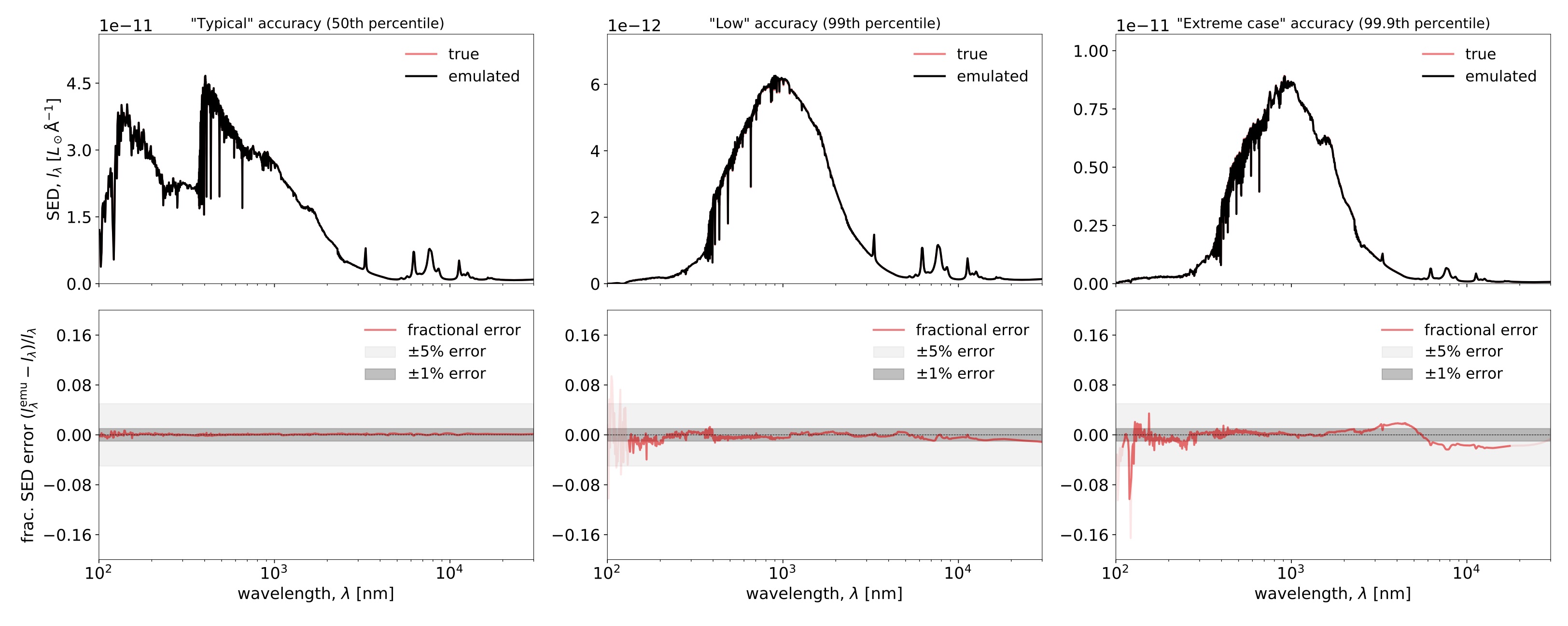}
\includegraphics[width = 17cm]{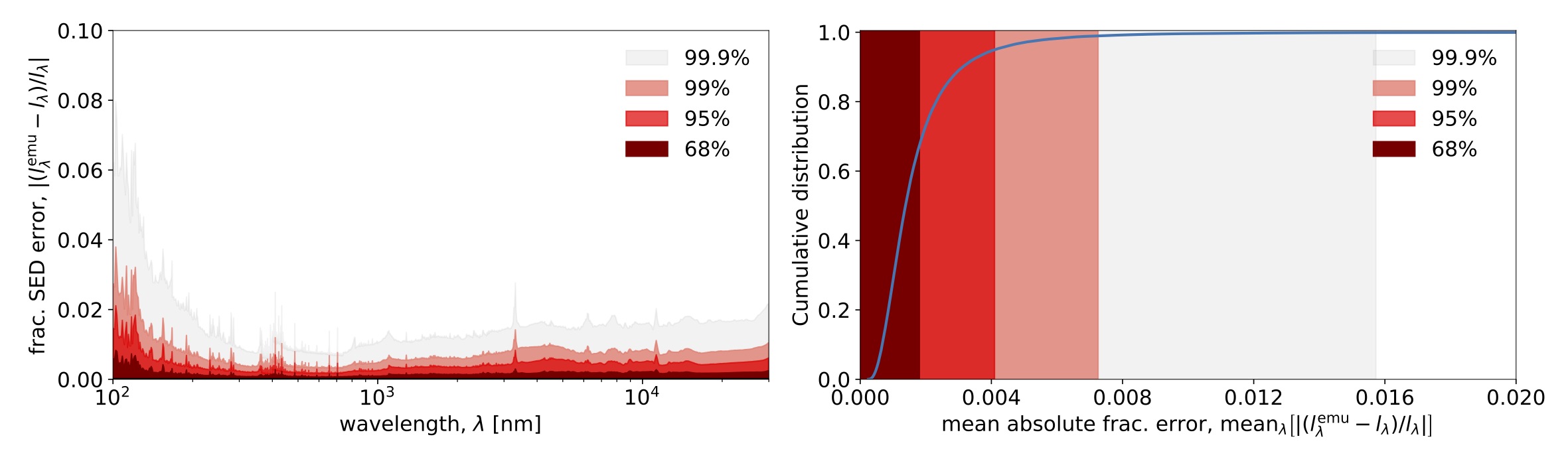}
\caption{Validation of the emulator for the Prospector-$\alpha$ model (\S \ref{sec:prospector}). Top figure: ``typical", ``low" and ``extreme case" accuracy for the emulated SEDs from a validation set of $10^5$ spectra generated with parameters drawn from the prior. These cases correspond to the 50th, 99th, and 99.9th percentiles of the mean (absolute) fractional error between the emulated and true SED (over the wavelength range). The displayed fractional errors (middle row) are faded out where the SEDs $\rightarrow0$. Bottom left: 68 (dark red), 95 (red), 99 (salmon) and 99.9 (grey) percentiles of the fractional emulator error as a function of wavelength. Bottom right: cumulative distribution and 68th (darkred), 95th (red), 99th (salmon) and 99.9th (grey) percentiles of the mean (absolute) fractional errors on the SEDs (over the full wavelength range). Typical errors (68\%) are sub-percent across the whole wavelength range. 99.9\% of samples are accurate to $< 2\%$ over most of the wavelength range, with the tails of the error distribution extending out to $\sim6\%$ at the shortest wavelengths.}
\label{fig:prospector_sed_accuracy}
\end{figure*}

Similarly to the DESI model, for validating the trained emulator we generated $10^5$ SEDs for model parameters drawn from the prior, and compared the emulated and exact SPS spectra for this validation set. The results are summarized in Figure \ref{fig:prospector_sed_accuracy}. The upper panels show typical, low and extreme case performance of the emulator, taken as the $50$th, $99$th, and $99.9$th percentiles of the mean (absolute) fractional error per SED (over the full wavelength range). The bottom left panel shows the $68$, $95$, $99$ and $99.9$ percent intervals of the fractional error as a function of wavelength, and the bottom right panel shows the cumulative distribution of the mean (absolute) fractional error for the validation samples (over the full wavelength range). Note that the emulator is trained on the PCA coefficients of log SEDs, but accuracy is shown in Figure \ref{fig:prospector_sed_accuracy} in linear space.

The emulator has typical fractional SED errors (68th percentile) at the $\ll 1\%$ level over the full wavelength range and parameter volume. $99.9\%$ of validation samples are accurate to better than $2\%$ down to $200\mathrm{nm}$, below which the accuracy steadily degrades with tails out to $\sim6\%$ at the lowest wavelengths (100$\mathrm{nm}$).
\subsection{Computational performance}
\label{sec:prospector_performance}
For the Prospector-$\alpha$ model, with the network architecture described in \S \ref{sec:prospector_emulator} the emulator is able to generate predicted SEDs a factor of $10^3$ faster (per wavelength range) than direct SPS computation with \textsc{fsps} on the same CPU architecture.

For applications where parallel SPS evaluations can be leveraged, the emulator can be called on a GPU without any additional development overhead. Generating $10^6$ emulated SEDs takes around $\sim 2\,\mathrm{s}$ on a Tesla K80 GPU, compared to $\sim 0.05\,\mathrm{s}$  per \textsc{fsps} call on an Intel i7 CPU; an overall factor of $10^4$ effective speed-up per SPS evaluation.

We leave investigation of additional performance gains enabled by the use of gradient based optimization and inference methods to future work.
\section{Validation III: Prospector-$\alpha$ photometry}
\label{sec:photometry}
In this section we demonstrate and validate direct emulation of photometry on the same Prospector-$\alpha$ model as considered in the previous section (see \S \ref{sec:prospector_model} and Table \ref{tab:prospector} for the model and parameters). 

\begin{figure*}
\centering
\includegraphics[width = 18cm]{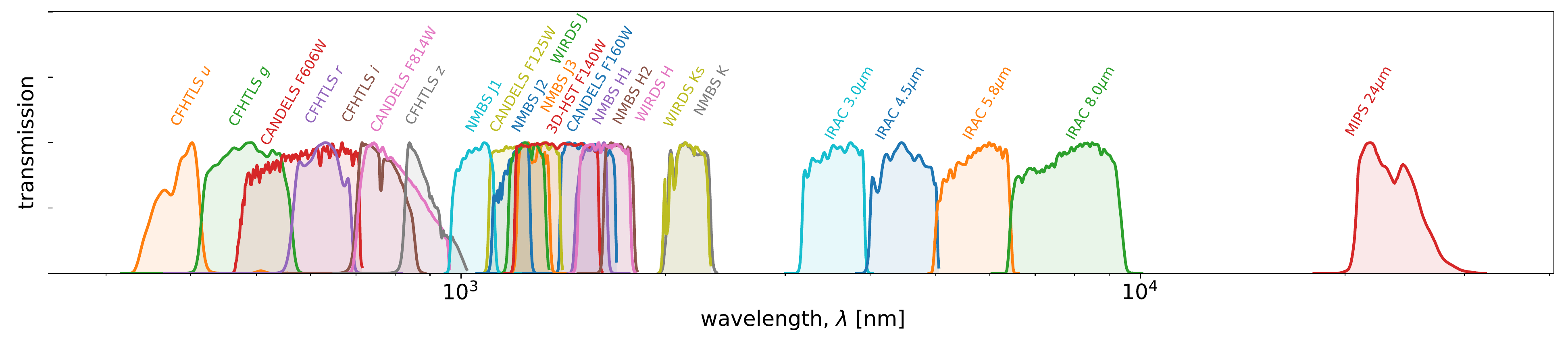}
\caption{The filters for the 24 bands emulated (for the Prospector-$\alpha$ model) in \S \ref{sec:photometry}, spanning the wavelength range $300\,\mathrm{nm}$ to $24\,\mu\mathrm{m}$.}
\label{fig:bands}
\end{figure*}

For this demonstration, we emulate the $24$ bands associated with the AEGIS field for the 3D-HST photometric catalog \citep{skelton2014}, supplemented by Spitzer/MIPS $24\mu m$ fluxes from \citep{whitaker2014}. This is motivated by the recent \citet{leja2019} analysis of the 3D-HST galaxies using the Prospector-$\alpha$ model. The 24 bands are as follows (shown in Figure \ref{fig:bands}): CFHTLS $ugriz$ \citep{erben2009}, CANDELS F606W, F814W, F125W, F160W \citep{grogin2011,Koekemoer2011}, NMBS J1, J2, J3, H1, H2, K \citep{whitaker2011}, WIRDS J, H, Ks \citep{bielby2012}, 3D-HST F140W \citep{brammer2012}, SEDS $3.6\mu$m and $4.5\mu$m \citep{ashby2013}, EGS $5.8\mu$m and $8.0 \mu$m \citep{barmby2008}, and Spitzer/MIPS $24\mu$m \citep{whitaker2014}.

In contrast to spectrum emulation in \S \ref{sec:prospector} where only rest-frame unit-mass SEDs were emulated (and mass and redshift adjusted afterwards as required), when emulating photometry we keep both mass $M$ and redshift $z$ as free parameters to be emulated over. Recall also that for photometry we will emulate the apparent magnitudes directly (\S \ref{sec:emu_phot}); there is no need for an intermediate (PCA) basis construction step in this case.
\subsection{Emulation}
We generated a training and validation set of $2\times 10^6$ and $1\times 10^5$ SEDs and associated photometry, for model parameters drawn from the prior (see Table \ref{tab:prospector}). We parameterized the apparent magnitudes for each band individually by a dense neural network with four hidden layers of 128 units each, with non-linear activation functions (Eq. \eqref{activation}) on all but the output layer, which has linear activation.

Network implementation and training follows exactly \S \ref{sec:fomo_emulator}.
\subsection{Results and validation}
\begin{figure*}
\centering
\includegraphics[width = 18cm]{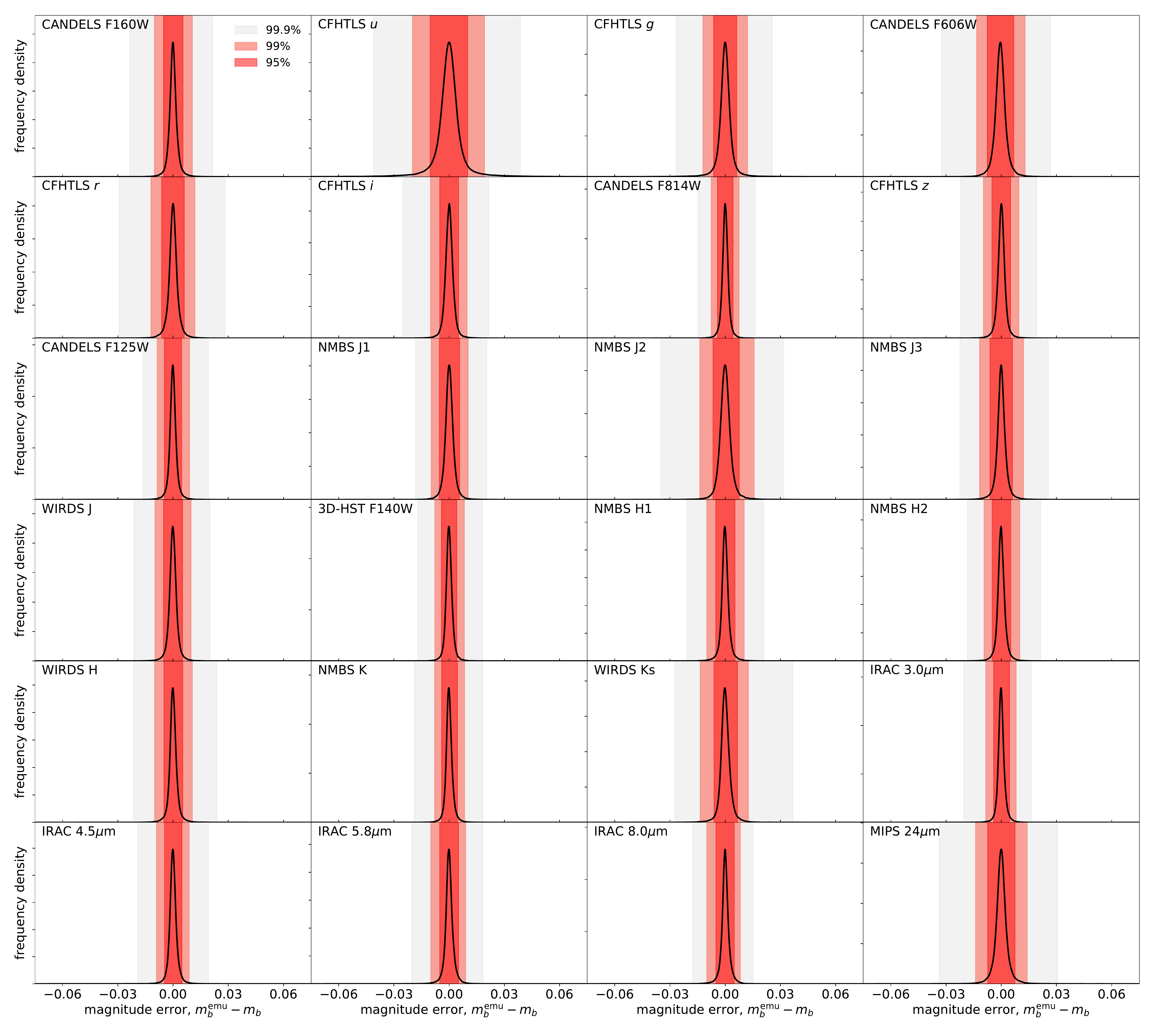}
\caption{Frequency densities (black) and 95 (red), 99 (salmon) and 99.9 (grey) percent intervals of the errors on the emulated apparent magnitudes for the 24 bands considered (\S \ref{sec:photometry}), over the $10^5$ samples in the validation set. For the chosen neural network architecture (\S \ref{sec:photometry}), the emulator is able to deliver percent-level accuracy across the board, with 99.9\% of validation samples being accurate to $\lesssim 0.02$ magnitudes for most bands, and $\lesssim 0.04$ in the worst cases.}
\label{fig:mag_accuracy}
\end{figure*}

The performance of the emulator is summarized in Figure \ref{fig:mag_accuracy}, which shows the frequency density (black) and 95 (red), 99 (salmon) and 99.9\% (grey) intervals of the emulator errors over the validation set, for all $24$ emulated bands. 
Across the board, the standard deviations of the error distributions are $<0.01$ magnitudes. For the majority of bands, 99.9\% of validation samples are accurate to better than $\lesssim 0.02$ magnitudes, and better than $\lesssim 0.04$ in the worst cases. In applications where an error floor of $0.05$ magnitudes is adopted due to SPS modeling and/or photometric calibration systematics, the emulator errors will make up a modest fraction of the total error budget.
\subsection{Computational performance}
We find that with the neural network architecture described above, the emulator is able to predict photometry a factor of $2\cdot 10^3$ faster (per band) than direct SPS computation for the Prospector-$\alpha$ model, with an additional order of magnitude speed-up when calling the emulator from the GPU. We find in experiments that larger network architectures give further improvements in accuracy, at the cost of some computational performance, and leave further optimization of network architectures for this problem to future work.
\section{Discussion and conclusions}
\label{sec:discussion}
SPS emulation offers a factor $\sim 10^3-10^4$ speed-up over direct SPS computation, whilst delivering percent-level accuracy over broad parameter and wavelength ranges. Parallel SPS evaluations can be further leveraged by calling the emulator from a GPU, giving an overall speed-up factor of $10^4-10^5$ compared to direct SPS evaluations on a CPU (for the models considered). In addition to the direct speed-up of SPS calls, the emulated SEDs and photometry come with readily accessible derivatives (with respect to the SPS model parameters), enabling the use of gradient-based inference and optimization methods; this is expected to reduce the number of SPS evaluations required when analyzing galaxy spectra or photometry under SPS models. The implications of the speed-up are clear: analyses that previously required significant high-performance computing investment could now be performed on a laptop, and previously intractable analyses of large populations of galaxies will now be tractable. For context, the $\sim1.5$ million CPU hour analysis of \citet{leja2019} could now be performed in $\sim$days on $16$-cores, and leveraging the gradients for inference is expected to give additional orders-of-magnitude improvement on top of that (e.g., \citealp{seljak2019}). Similarly, the computational cost associated with SPS evaluation when forward-simulating large surveys will be radically reduced.

Whilst the specific SPS models presented in this paper were motivated by analysis of photometry and low S/N spectra respectively, another promising area for emulation is SPS models designed to fit high S/N, high resolution galaxy spectra. These models are often computationally expensive ($\sim$1 minute per SPS evaluation) and are thus particularly attractive candidates for speed-up by emulation. However, the model dimensionality and required precision can be demanding. For the simple case of quiescent galaxies, state-of-the-art models have up to $\sim$40 parameters which control components such as the initial mass function, individual elemental abundances, as well as detailed models of continuum line spread functions (e.g., \citealt{conroy12,conroy18}). The systematic residuals for such models are on the order of 1\%, so in practice an emulator would need to reproduce thousands of pixels to sub-percent-level accuracy. Star-forming galaxies bring additional challenges, notably nebular emission -- photoionisation codes can have hundreds of parameters controlling hundreds of emission lines \citep{ferland17}, of which each emission line in principle could have its own line spread function. Although the model complexity and fidelity requirements are higher for this use case, because these models are so much more expensive one has considerably more leeway in using larger and more sophisticated neural network architectures, whilst still potentially achieving significant computational speed-up.

Another avenue that SPS emulation opens up is Bayesian hierarchical analysis of large galaxy populations under SPS models, i.e., jointly inferring the physical properties of individual galaxies in a sample along with the intrinsic (prior) distribution of galaxy characteristics. The high-dimensional inference tasks associated with such analyses typically requires gradient-based inference algorithms, such as Hamiltonian Monte Carlo sampling, which will be made substantially easier with emulated SPS.

There are a number of areas where the neural network emulation framework presented here can be improved upon. Firstly, we did not go to great lengths to optimize the neural network architectures to deliver the optimal trade-off between accuracy and speed-up. Once the training sets have been generated, training the emulator networks is sufficiently cheap that a search over network architectures (including more sophisticated architecture types) to deliver the best performance is computationally feasible. 

Regarding basis construction for galaxy spectra, we have shown that PCA is effective for a range of applications. However, for complex SPS models or where fidelity requirements are very high, alternative basis constructions such as non-negative matrix factorization (NMF) in linear flux \citep{hurley2014,lovell2019}, or non-linear representation construction with autoencoders, may prove more powerful.

The other area where some additional effort could give substantial improvements is intelligently sampling the parameter space when building the training set. In this study, little focus was given to optimizing parameter space sampling and training set size; training set sizes were simply chosen so that they could be generated in $\lesssim$ days and deliver percent-level accuracy in the trained emulators. However, it is clearly advantageous to use online learning to optimally sample the parameter space on-the-fly in conjunction with the emulator training (see e.g., \citealp{rogers2019,alsing2019}). This approach has the benefits that it both enables more optimal sampling of the parameter space, and by generating the training set synchronously with training, the size of the training set required to achieve a given accuracy target can be determined on-the-fly (i.e., training and acquisition of training data can be stopped when the accuracy reaches the desired threshold).

For inference applications when the emulator error cannot safely be assumed to be a negligible fraction of the total error budget, it will be desirable to have some quantification of the emulator uncertainties that can be folded into the likelihood function. This can be achieved within the neural network paradigm by using Bayesian neural networks: performing posterior inference over the network weights given the training data (and some priors), hence providing posterior predictive distributions over the output SEDs or photometry rather than simple point estimates. This sophistication comes at the cost of having to perform many forward passes through the network to obtain an emulator error estimate at a given set of SPS parameters.

The emulation code -- \textsc{speculator} -- is publicly available at \url{https://github.com/justinalsing/speculator}.
%
%\section{Conclusions}
%\label{sec:conclusions}
%Computing stellar population synthesis models of galaxy spectra is the main computational bottleneck for forward-modeling large surveys and inferring physical parameters from observed spectra or photometry. We have developed a simple framework for fast emulation of SPS model spectra, using principal component analysis to construct a set of basis functions, and neural networks to learn the basis coefficients as a function of the SPS model parameters. For photometry, we parameterize the magnitudes as a function of SPS parameters as a neural network. The resulting emulators provide compact, differentiable representations of SPS models that are both fast and accurate. 
%
%We have validated the emulator on two SPS models -- a relatively simple eight-parameter model, and a more flexible 14-parameter model -- and demonstrated that the emulator can deliver a factor $\sim 10^3-10^4$ speed-up over direct SPS computation whilst providing percent-level precision over broad parameter prior and wavelength ranges. Calling the emulator from a GPU is straightforward, giving an additional order-of-magnitude speed-up compared to direct SPS computation on a CPU. Furthermore, with readily-available derivatives, the emulator will allow for use of gradient-based optimization and inference methods, reducing the number of SPS model evaluations required to make robust inferences from data. We anticipate that rapid SPS emulation will be of central importance for future analysis and simulation of galaxy populations under SPS models.
%
\acknowledgments
We thank Benjamin Joachimi and Fran\c{c}ois Lanusse for useful discussions. JA and HVP were partially supported by the research project grant ``Fundamental Physics from Cosmological Surveys" funded by the Swedish Research Council (VR) under Dnr 2017-04212. HVP, BL and DJM acknowledge the hospitality of the Aspen Center for Physics, which is supported by National Science Foundation grant PHY-1607611. JL is supported by an NSF Astronomy and Astrophysics Postdoctoral Fellowship under award AST-1701487. This work was also partially supported by a grant from the Simons Foundation, and partially enabled by funding from the UCL Cosmoparticle Initiative.
%\appendix
%\section{Appendix 1}

\bibliography{emulator}

\end{document}